\documentclass[aps,superscriptaddress,showpacs,preprint]{revtex4}
\usepackage{graphics}
\begin{document}
\title{Anomalous bias dependence of tunnel magnetoresistance in a magnetic 
tunnel junction}
\author{Soumik Mukhopadhyay}
\author{I. Das}
\affiliation{ Saha Institute of Nuclear Physics,1/AF,Bidhannagar,
Kolkata 700064, India}
\author{S. P. Pai}
\author{P. Raychaudhuri}
\affiliation{ 
Tata Institute of Fundamental Research, 
Homi Bhabha Road, Mumbai 400005, India\\
(Received
}

\begin{abstract}
We have fabricated a spin-polarized tunneling device based 
on half metallic manganites incorporating $Ba_{2}LaNbO_{6}$ 
as insulating barrier. An anomalous bias dependence of 
tunnel magnetoresistance (TMR) has been observed, the first of its kind 
in a symmetric electrode tunnel junction with single insulating barrier. 
The bias dependence of TMR shows an extremely sharp zero bias anomaly, 
which can be considered as a demonstration of the drastic 
density of states variation around the Fermi level of the half metal. This  
serves as a strong evidence for the existence 
of minority spin tunneling states at the half-metal insulator interface.
\end{abstract}
\maketitle
\begin{figure}
\resizebox{8cm}{6cm}
{\includegraphics {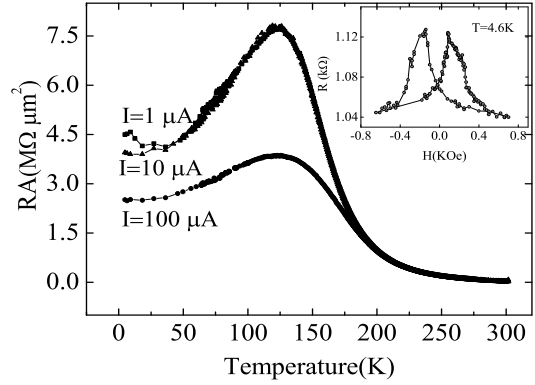}}
\caption{Junction resistance-area product(RA) at
different bias currents in absence of magnetic field.
Junction resistance shows weaker temperature
dependence at higher bias current. Inset: Junction resistance vs. magnetic
field at 4.6K}\label{fig:res}
\end{figure}
\begin{figure}
\resizebox{8cm}{6cm}
{\includegraphics{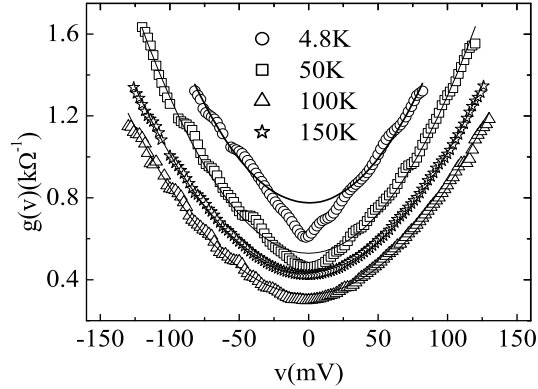}}
\caption{Differential junction conductance vs. voltage curves at
different temperatures in absence of magnetic field. The continuous lines are
the Brinkman fits.}\label{fig:cond}
\end{figure}
\begin{figure}
\resizebox{8cm}{6cm}
{\includegraphics{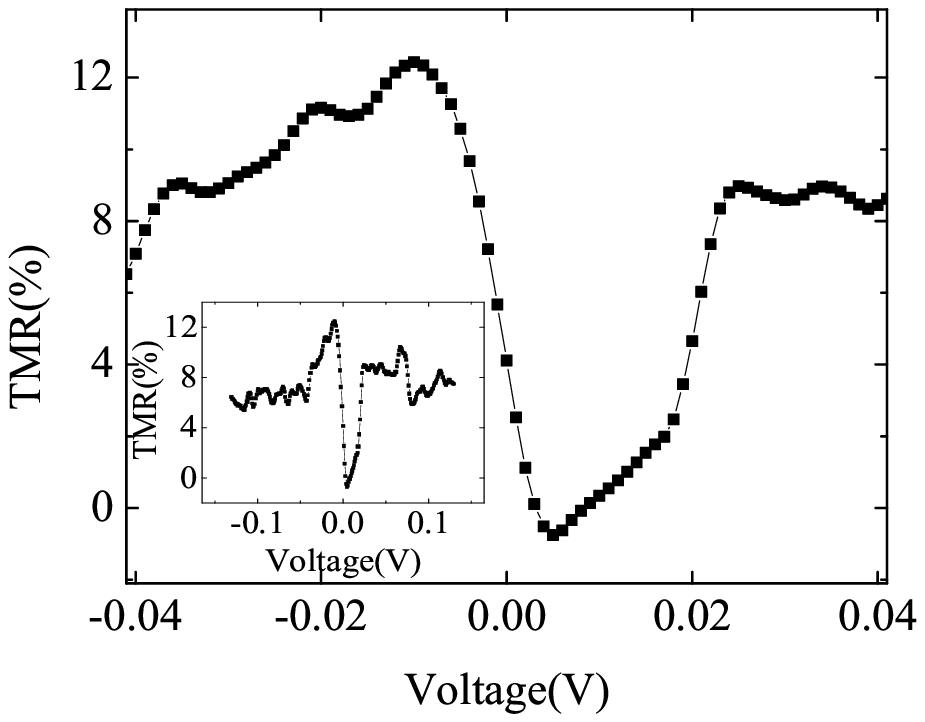}}
\caption{Bias dependence of TMR at 6K highlighting the anomalous behaviour around
zero bias. Inset: Bias dependence in the experimental voltage range}\label{fig:bias}
\end{figure}
\begin{figure}
\resizebox{5cm}{4cm}
{\includegraphics{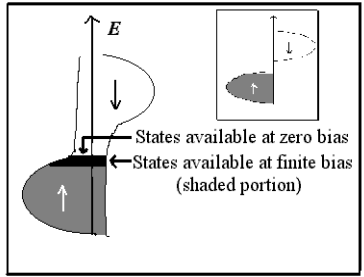}}
\caption{Modified band structure of LSMO at the interface showing minority states
around the Fermi level and the majority spin DOS increasing with increasing bias. Inset:
Band structure of bulk LSMO}\label{fig:band}
\end{figure}

\indent
In recent years Magnetic Tunnel Junction (MTJ)~\cite{jag} has become 
the subject of great interest because of the richness in physical 
properties it exhibits, and potential applications. 
It has been observed that the transport properties of MTJ depend not only on the 
ferromagnetic metal electrodes but also on the insulator-electrode couple and 
its interface and that the insulator plays a crucial role 
in selecting bands that can tunnel~\cite{desci,sharma}. 
However, not much thought has been given 
to what happens to the band structure at the interface 
between the half metal and the insulator in MTJs consisting of half metals. 
Till date there is no evidence of minority spin states in manganite tunnel junction
interfaces, although there are reports of existence of minority 
spin states in manganites~\cite{minority}. 
The present article establishes the presence of \emph{minority spin tunneling
states} in half metal near the half metal-insulator interfaces. 
An extremely sharp zero bias anomaly in the bias dependence of 
TMR has been observed.\\
\indent
In almost all MTJs, the Tunnel Magnetoresistance
(TMR; $\delta R/R = (R_{AP}-R_{P})/R_{AP}$ where $R_{AP}$, $R_{P}$ are the resistances
in the antiparallel and parallel magnetization configuration respectively.)
decreases with increasing bias voltage~\cite{jag}. Several mechanisms 
have been proposed to explain this behaviour. For example, there are processes 
like spin independent two-step elastic tunneling via defect states~\cite{defect} 
in the insulating barrier and inelastic processes like spin flip scattering by 
magnon excitations~\cite{mag1,mag2}, magnetic impurities~\cite{impure}
in the barrier etc. The effect of density of states (DOS)~\cite{dos} 
or the influence of electric field on the barrier height is 
often discussed in explaining the variation of TMR at higher bias. 
It is believed that in the low bias region the electronic band 
structure of the electrodes play fundamental roles. Yet we do not observe the 
effect of sharp variation of electronic density of states (DOS) about the 
Fermi level at zero bias (for reasons to be discussed later). This report is
a clear demonstration of the effect of drastic DOS variation around 
the Fermi level near the interface.\\
\indent
Here we report the transport properties of a manganite 
MTJ where $Ba_{2}LaNbO_{6}$ has been introduced as the 
insulating barrier. The MTJ was prepared by 
pulsed laser deposition. The trilayer $La_{0.67}Sr_{0.33}MnO_{3}$
(LSMO)/$Ba_{2}LaNbO_{6}$(BLNO)/LSMO 
was deposited on single crystalline $SrTiO_{3}(100)$ substrate held at a 
temperature $800^{0}C$ and the oxygen pressure was 400 mTorr. 
BLNO has a complex cubic perovskites structure and can be grown epitaxially 
on single crystal perovskite substrates~\cite{blno1,blno2}. The thickness 
of the bottom LSMO layer is $1000\AA$  and that of the top layer $500\AA$ while the 
estimated thickness of the insulating spacer from the deposition 
rate calibration of BLNO is $50\AA$. The microfabrication in the cross-strip geometry 
was done using photolithography and ion-beam milling. The bottom layer was
patterned using photolithography while the top electrode and the insulating layer
using ion-beam milling. The junction area 
is $50 \times 50 {\mu}m^{2}$. The transport and magnetoresistive 
properties were measured in the current perpendicular to plane geometry 
using four terminals method with the magnetic field applied in the 
plane of the sample.\\
\indent
The junction resistance in the absence of magnetic field shows a distinct
peak at around 125 K
(Fig:~\ref{fig:res}), typical of manganite tunnel junctions~\cite{res,res1}.
With increase in bias level the temperature dependence of junction 
resistance becomes weaker(Fig:~\ref{fig:res}). 
The conductance curves show parabolic voltage dependence(Fig:~\ref{fig:cond}). 
We have fitted the differential conductance vs. voltage curves
using asymmetric barrier Brinkman model~\cite{bri} in different voltage ranges. The
average barrier height and the barrier width turns out to be in the
range $0.2- 0.25$ eV and $32-37\AA$ respectively.
The asymmetry in the barrier obtained from the Brinkman model is very small,
about $3-4$ mV only and hence the current-voltage characteristics can be well fitted
with symmetric barrier Simmons model~\cite{sim}, producing similar results.
The barrier parameters like average barrier height and barrier width are almost
temperature independent within the relevant temperature range.
All these observations indicate that the device
is free of any pinhole shorts and tunneling is the dominant
transport mechanism~\cite{short}.
The highest value of TMR obtained at any bias current is 
around $10\%$ . Low TMR value signifies a considerable 
reduction of spin polarization at the electrode-barrier interface. The 
tunnel magnetoresistance vanishes above 150K, as is the case generally for manganite 
tunnel junctions~\cite{temp}.\\
\indent
The observed bias dependence of TMR(Fig:~\ref{fig:bias}) has an unusual feature. 
Previous reports on bias dependence show that Tunnel 
Magnetoresistance decreases with increasing bias voltage and there is not much
appreciable variation around zero bias~\cite{jag}. Here we have observed 
that the TMR undergoes a sharp rise(Fig:~\ref{fig:bias}) with increasing 
bias voltage 
below $20-25$ mV at $6$K. Only one or two groups~\cite{desci,sharma,suzuki} have seen 
bias dependence like this but there are some striking differences
between their observations and ours. For example, 
De Teresa et al.~\cite{desci} observed such 
bias dependence in case of $Co/SrTiO_{3}/La_{0.67}Sr_{0.33}MnO_{3}$ hybrid 
tunnel Junction and Sharma et al.~\cite{sharma} observed this in case of 
MTJ with composite 
barrier where the TMR undergoes a sharp decrease as one approaches 
zero bias. In all those cases highly asymmetric barrier is the key factor
in the anomalous bias dependece of TMR.
But here we are dealing with a half metallic tunnel 
junction with single insulating barrier. There is a marginal asymmetry
in the bias dependence. This is due to a small barrier asymmetry 
close to $3-4$ mV which is very small compared to the 
average barrier height of $0.20-0.25$ eV. This is manifested in the
conductance minimum at low temperature(Fig:~\ref{fig:cond}) being shifted with 
respect to zero bias.\\
\indent
There are several factors contributing to the bias dependence of TMR. If there 
exists defect sites within the barriers, creation of states either thermally
or by hot electron impact will facilitate two-step tunneling. Since 
these states are not polarized, the two-step tunneling is spin independent and
does not contribute to TMR. With increase in bias voltage, density of available
defect states increases exponentially. As a result the two-step tunneling 
current increases 
sharply with increasing bias voltage, thus reducing the TMR with increasing 
bias. There are other inelastic processes that can infuence the bias 
dependence. For example ``hot electrons'' tunneling across the insulating 
barrier may lose their energy by emitting a magnon and thereby flipping the
electron spin. With increasing bias more magnons can be emitted, resulting in
reduced TMR. Spin flip scattering cross-section due to magnetic impurities
in the barrier increases with increasing bias thus reducing the TMR. The
existence of coulomb gap due to metallic inclusions at the junction interface
also reduces the TMR with increasing bias~\cite{res}. Obviously 
these higher order tunneling processes cannot explain
the anomalous bias dependence around zero bias.\\
\indent
Biasing an MTJ leads to the contributions from electrons, which tunnel from the 
occupied states below the Fermi level of one electrode to the empty states at 
the same energy above the fermi level of the other electrode. Due to the 
change in DOS of ferromagnets as a function of energy, the spin polarization 
should be voltage dependent and hence the TMR. But in practice, 
the nature of bias dependence in almost all MTJ's does not seem to reflect that.
This is because for transition metals, although most of the spin polarization
comes from d-band, majority of the tunneling electrons are from s-band which 
is not sharply polarized. There is no significant change of this situation if 
we take into account the effect of s-d hybridization~\cite{sd}. On the other 
hand, in half metals like manganites, 
the spin up and spin down bands are completely split (Fig:~\ref{fig:band}) 
resulting in the spin polarization being immuned to the sharp variation in DOS near 
the band edge. Thus the DOS variation with change in bias voltage does not affect the
spin polarization.\\
\indent
The observed bias dependence can be interpreted as follows. Due to the  
perturbation caused by the complex band structure of the  
insulating barrier, the spin split band edges at the 
interfaces get modified in such a way that there is a substantial overlap
of the up and down spin bands below the Fermi level(Fig:~\ref{fig:band}). 
This results in the 
introduction of minority spin states. The existence of minority spin 
states will bring the variation in majority spin DOS into play 
because then it will strongly influence the tunneling spin polarization.
At low bias, the DOS 
for the majority spin band around Fermi level is much less than at higher 
voltage and the DOS slope is also much sharper at near the Fermi level. 
With increase in bias voltage from zero value, the majority spin DOS increases 
sharply (Fig:~\ref{fig:band}).
Hence within a certain bias window around zero bias the spin polarization 
can have a sharp rise with increasing bias. 
Still one should expect the TMR to increase beyond the observed
bias range considering the large band width in LSMO (half band width $0.75eV$). 
But at higher votages the opening up of inelastic conduction channel is bound to 
have an influence on the bias dependence of TMR. Thus we get the fingerprint of the 
DOS profile only in the low bias range.\\
\indent
A small inverse TMR (where $R_{AP} < R_{P}$) is also observed in the extreme 
vicinity of zero bias. Inverse TMR is expected in systems with the two 
different electrodes having opposite spin polarizations within a certain 
bias range~\cite{pratap}. In our case the majority and minority spin DOS
become comparable near the Fermi level at zero bias. And since there is 
a small relative zero bias shift of Fermi levels between the two 
electrodes, it may give rise to Inverse TMR near zero bias.\\
\indent
To summarize, we propose that the observed anomalous bias dependence of TMR 
can be considered as the 
{\emph{experimental demonstration of the influence of interfacial DOS 
around zero bias in a symmetric MTJ}} 
and that the influence of the insulating barrier can lead to the 
formation of {\emph{minority spin tunneling DOS}} at the interfaces of 
a half metallic tunnel junction.

\end{document}